\documentclass{llncs}
\usepackage{graphicx}
\usepackage{amssymb}
\usepackage{amsmath}
\usepackage{amsfonts}
\usepackage{algorithm}
\usepackage{bm}
\usepackage{pdflscape}

\newcommand{\vs}{\vspace{.5ex}}
\newcommand{\ds}{\displaystyle}
\begin{document}
	\title{Improved Job Sequencing Bounds \\from Decision Diagrams}
	%
	%
	\author{J. N. Hooker}
	\authorrunning{J. N. Hooker}
	%
	\institute{Carnegie Mellon University\\
		\email{jh38@andrew.cmu.edu}}
	\maketitle              
	\begin{abstract}
		We introduce a general method for relaxing decision diagrams that allows one to bound job sequencing problems by solving a Lagrangian dual problem on a relaxed diagram.  We also provide guidelines for identifying problems for which this approach can result in useful bounds.  These same guidelines can be applied to bounding deterministic dynamic programming problems in general, since decision diagrams rely on DP formulations.  Computational tests show that \mbox{Lagrangian} relaxation on a decision diagram can yield very tight bounds for certain classes of hard job sequencing problems.  For example, it proves for the first time that the best known solutions for Biskup-Feldman instances are within a small fraction of 1\% of the optimal value, and sometimes optimal. 
	
	\end{abstract}

	\section{Introduction}
	
	In recent years, binary and multivalued decision diagrams (DDs) have emerged as a useful tool for solving discrete optimization problems \cite{BerCirHoeHoo14,BerCirHoeHoo16,SerHoo19}.  
	A key factor in their success has been the development of {\em relaxed} DDs, which represent a superset of the feasible solutions of a problem and provide a bound on its optimal value.  While an exact DD representation of a problem tends to grow exponentially with the size of the problem instance, a relaxed DD can be much more compact when properly constructed.  The tightness of the relaxation can be controlled by adjusting the maximum allowed width of the DD.  
	
	
	Relaxed DDs are normally used in conjunction with a branching procedure \cite{BerCirHoeHoo14,CirHoe13}, much as is the linear programming (LP) relaxation in an integer programming solver.  As branching proceeds, the relaxed diagram provides a progressively tighter bound.  However, combinatorial problems are often solved with heuristic methods that do not involve branching.  This is true, in particular, of job sequencing problems.  In such cases it is very useful to have an independently derived lower bound that can provide an indication of the quality of the solution.  
	
	Recent research \cite{Hoo17} has found that a relaxed DD can yield good bounds for hard job sequencing problems without branching.  In fact, a surprisingly small relaxed DD, generally less than 10\% the width of an exact DD, can yield a bound equal to the optimal value. 	On the other hand, since exact DDs grow rapidly with the instance size, relaxed DDs that are 10\% of their width likewise grow rapidly.  As a result, relaxed DDs of reasonable width tend to provide progressively weaker bounds as the instances scale up.
	
	It is suggested in \cite{Hoo17} that Lagrangian relaxation could help strengthen the bounds obtained from smaller relaxed DDs.  In this paper, we propose a general technique for relaxing a DD while preserving the ability to obtain  Lagrangian bounds from the DD.  The relaxed DD is constructed by merging nodes only when they agree on certain state variables that are crucial to forming the Lagrangian relaxation.  
	
	We find that for certain types of job sequencing problems,  Lagrangian relaxation in relaxed DDs of reasonable width can provide very tight bounds on the optimal value.  For example, we prove for the first time that the best known solution values of Biskup-Feldman single-machine scheduling instances are within a small fraction of one percent of the optimum, and sometimes optimal.  
	
Furthermore, we identify general conditions under which Lagrangian relaxation can be implemented in a relaxed DD for purposes of obtaining bounds.  The conditions are expressed in terms of structural characteristics of the \mbox{dynamic} programming model that defines the DD.  They lead to a new tool for bounding not only job sequencing problems with suitable structure, but general deterministic dynamic programming models that satisfy the conditions.

	\section{Previous Work}
	
	Decision diagrams were introduced as an optimization method by \cite{HadHoo07,Hoo06d}.  The idea of a relaxed diagram first appears in \cite{AndHadHooTie07} as a means of enhancing propagation in constraint programming.  Relaxed DDs were first used to obtain optimization bounds in \cite{BerCirHoeHoo13,BerHoeHoo11}.  Connections between DDs and deterministic dynamic programming are discussed in \cite{Hoo13}.  
	
	Bergman, Cir\'{e} and van Hoeve first applied Lagrangian relaxation to decision diagrams in \cite{BerCirHoe15}, where they use it successfully to strengthen bounds for the traveling salesman problem with time windows.  They also use Lagrangian relaxation and DDs in \cite{BerCirHoe15a} to improve constraint propagation.  
	
	We advance beyond Bergman et al.\ \cite{BerCirHoe15} in two ways.  First, we show how to obtain bounds on tardiness and a variety of other objective functions from a stand-alone relaxed DD.  The DD in \cite{BerCirHoe15} represents only an all-different constraint and can provide bounds only on total travel time (without taking time windows into account).  The DD is embedded in a constraint programming (CP) model that contains the time window constraints.
	While constraints could be added to the CP model to obtain tardiness and other kinds of bounds from the CP solver, the DD itself cannot provide them. One or more additional state variables are necessary, which results in a more complicated DD than the one used in \cite{BerCirHoe15}.  Our contribution is to define a new node merger scheme that relaxes such a DD while allowing Lagrangian relaxation to be applied.
	
Our second contribution is to analyze, in general, when and how Lagrangian \mbox{relaxation} can be combined with DDs. We introduce the concepts of an exact state and an immediate penalty function and use these concepts to formulate sufficient conditions for implementing Lagrangian relaxation in a relaxed DD. This leads to a general method for bounding dynamic programming models that satisfy the conditions.  We find that while the method generates impracticably large relaxed DDs for the job sequencing problems in \cite{BerCirHoe15} and \cite{Hoo17}, it is quite practical for several important types of job sequencing problems.

	\section{Decision Diagrams}
	
	For our purposes, a decision diagram can be defined as a directed, acyclic multigraph in which the nodes are partitioned into {\em layers}.  Each arc of the graph is directed from a node in layer $i$ to a node in layer $i+1$ for some $i\in \{1,\ldots, n\}$.  Layers 1 and $n+1$ contain a single node, namely the root $r$ and the terminus $t$, respectively.  Each layer $i$ is associated with a finite-domain variable $x_i\in D_i$.  The arcs leaving any node in layer $i$ have distinct {\em labels} in $D_i$, representing possible values of $x_i$ at that node.  A path from  $r$ to $t$ defines an assignment to the tuple $x=(x_1,\ldots,x_n)$ as indicated by the arc labels on the path.  The decision diagram is {\em weighted} if there is a length (cost) associated with each arc.  
	
	Any discrete optimization problem with finite-domain variables can be represented by a weighted decision diagram.
	The diagram is constructed so that its $r$--$t$ paths correspond to the feasible solutions of the problem, and the length (cost) of any $r$--$t$ path is the objective function value of the corresponding solution.  If the objective is to minimize, the optimal value is the length of a shortest $r$--$t$ path.  Many different diagrams can represent the same problem, but for a given variable ordering, there is a unique {\em reduced} diagram that represents it \cite{Bry86,Hoo13}.  
	
\begin{figure}[!t]
		\centering
		\includegraphics[clip=true,trim=150 480 100 120,scale=.85]{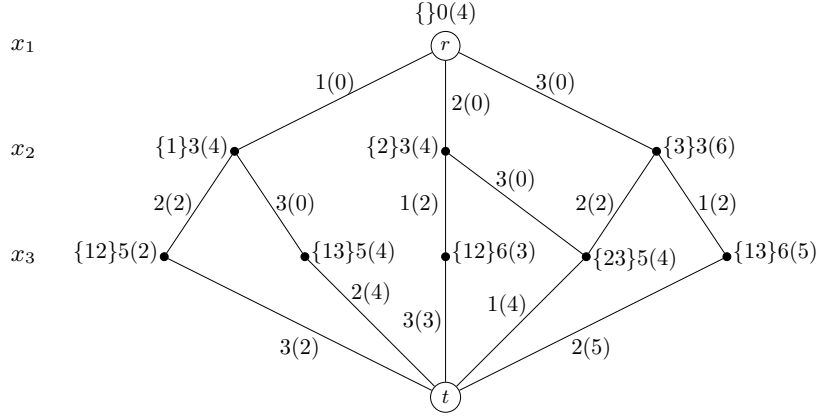}
		\vspace{-3ex}
		\caption{Decision diagram for a small job sequencing instance, with arc labels and costs shown.  States and minimum costs-to-go are indicated at nodes.}   \label{fig:DDstates}
	\end{figure}

	As an example, consider a job sequencing problem with time windows.  Each job $j$ begins processing no earlier than the release time $r_j$ and requires processing time $p_j$.  The objective is to minimize total tardiness, where the tardiness of job $j$ is $\max\{0,s_j+p_j-d_j\}$, and $d_j$ is the job's due date.  Figure~\ref{fig:DDstates} shows a reduced decision diagram for a problem instance with $(r_1,r_2,r_3)=(0,1,1)$, $(p_1,p_2,p_3)=(3,2,2)$, and $(d_1,d_2,d_3)=(5,3,5)$.  Variable $x_i$ represents the $i$th job in the sequence, and arc costs appear in parentheses.

	\section{Dynamic Programming Models}
	
	Decicision diagrams most naturally represent problems with a dynamic programming formulation, because in this case a simple top-down compilation procedure yields a DD that represents the problem.  A general dynamic programming formulation can be written
	\begin{equation}
	{\ds
		h_i(\bm{S}_i) = \min_{x_i\in X_i(\bm{S}_i)} \Big\{ c_i(\bm{S}_i,x_i) + h_{i+1}\big(\phi_i(\bm{S}_i,x_i)\big)\Big\}
	}
	\label{eq:recursion}
	\end{equation}
	Here, $\bm{S}_i$ is the {\em state} in stage $i$ of the recursion.  Typically the state is a tuple $\bm{S}_i=(S_{i1},\ldots,S_{ik})$ of {\em state variables}.  Also $X_i(\bm{S}_i)$ is the set of possible {\em controls} (values of $x_i$) in state $\bm{S}_i$, $\phi_i$ is the {\em transition function} in stage $i$, and $c_i(\bm{S}_i,x_i)$ is the {\em immediate cost} of control $x_i$ in state $\bm{S}_i$.  We assume there is single  initial state $\bm{S}_1$ and a single final state $\bm{S}_{n+1}$, so that $h_{n+1}(\bm{S}_{n+1})=0$ and $\phi_n(\bm{S}_n,x_n)=\bm{S}_{n+1}$ for all states $\bm{S}_n$ and controls $x_n\in X_n(S_n)$.  The quantity $h_i(\bm{S}_i)$ is the {\em cost-to-go} for state $\bm{S}_i$ in stage $i$, and an optimal solution has value $h_1(\bm{S}_1)$.
	
	In the job sequencing problem, the state $\bm{S}_i$ is the tuple $(V_i,t_i)$, where state variable $V_i$ is the set of jobs scheduled so far, and state variable $t_i$ is the finish time of the last job scheduled.  Thus the initial state is $\bm{S}_1=(\emptyset,0)$, and  $X_i(S_i)$ is $\{1,\ldots,n\}\setminus V_i$.  The transition function $\phi_i(\bm{S}_i,x_i)$ is given by
	\[
	\phi_i\big((V_i,t_i),x_i\big) = \big(V_i\cup\{x_i\},\;\max\{r_{x_i},t_i\}+p_{x_i}\big)
	\]
	The immediate cost is the tardiness that results from scheduling job $x_i$ in state $(V_i,t_i)$.  Thus if $\alpha^+ = \max\{0,\alpha\}$, we have
	\begin{equation}
	c_i\big((V_i,t_i),x_i\big) = \Big(\max\{r_{x_i},t_i\}+p_{x_i}-d_{x_i}\Big)^+
	\label{eq:dp5}
	\end{equation}

	We recursively construct a decision diagram $D$ for the problem by associating a state with each node of $D$.  The initial state $\bm{S}_1$ is associated with the root node $t$ and the final state $\bm{S}_{n+1}$ with the terminal node $t$.  If state $\bm{S}_i$ is associated with node $u$ in layer $i$, then for each $v_i\in X_j(\bm{S}_i)$ we generate an arc with label $v_i$ leaving $u$.  The arc terminates at a node associated with state $\phi_i(\bm{S}_i,v_i)$.  Nodes on a given layer are identified when they are associated with the same state.  
	
	The process is illustrated for the job sequencing example in Fig.~\ref{fig:DDstates}.  Each node is labeled by its state $(V_i,t_i)$, followed (in parentheses) by the minimum cost-to-go at the node.  The cost-to-go at the terminus $t$ is zero.  
	

	\section{Relaxed Decision Diagrams}
	
	A weighted decision diagram $D'$ is a {\em relaxation} of diagram $D$ when $D'$ represents every solution in $D$ with equal or smaller cost, and perhaps other solutions as well.  To make this more precise, suppose layers $1,\ldots,n$ of both $D$ and $D'$ correspond to variables $x_1,\ldots,x_n$ with domains $X_1,\ldots,X_n$.  Then $D'$ is a relaxation of $D$ if every assignment to $x$ represented by an $r$--$t$ path $P$ in $D$ is represented by an $r$--$t$ path in $D'$ with length no greater than that of $P$.  The shortest path length in $D'$ is a lower bound on the optimal value of the problem represented by $D$.  We will refer to a diagram that has not been relaxed as {\em exact}.
	
	We can construct a relaxed decision diagram in top-down compilation by {\em merging} some nodes that are associated with different states.  The object is to limit the width of the diagram (the maximum number of nodes in a layer).  When we merge nodes with states $\bm{S}$ and $\bm{T}$, we associate a state $\bm{S}\oplus \bm{T}$ with the resulting node.  The operator $\oplus$ is chosen so as to yield a valid relaxation of the given recursion.  
	

	\begin{figure}[!b]
		\centering
		\includegraphics[clip=true,trim=150 480 60 120,scale=.85]{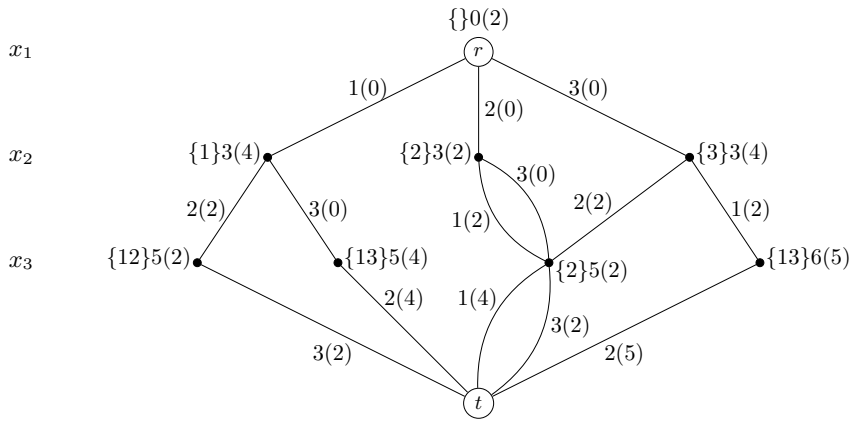}
		\vspace{-1ex}
		\caption{A relaxation of the decision diagram in Fig.~\ref{fig:DDstates}.}
		\label{fig:DDrelax}
	\end{figure}

	The job sequencing problem discussed above uses a relaxation operator
	\[
	(V_i,t_i)\oplus (V'_i,t'_i)=(V_i\cap V'_i,\min\{t_i,t'_i\})
	\]  
	$V_i$ is now the set of jobs scheduled along all paths to the current node, and $t_i$ is the earliest finish time of the last scheduled jobs along these paths.  The operator is illustrated in Fig.~\ref{fig:DDrelax}, which is the result of merging states $(\{1,2\},6)$ and $(\{2,3\},5)$ in layer~3 of Fig.~\ref{fig:DDstates}.  The relaxed states $(V,f)$ are shown at each node, followed by the minimum cost-to-go in parentheses.  The shortest path now has cost 2, which is a lower bound on the optimal cost of 4 in Fig.~\ref{fig:DDstates}.
	
	Sufficient conditions under which node merger results in a relaxed decision diagram are developed in \cite{Hoo17}.  A state $\bm{S}'_i$ {\em relaxes} a state $\bm{S_i}$ when (a) all feasible controls in state $\bm{S_i}$ are feasible in state $\bm{S}'_i$, and (b) the immediate cost of any given feasible control in $\bm{S}_i$ is no less than its immediate cost in $\bm{S}'_i$.  That is, $X_i(\bm{S}_i)\subseteq X_i(\bm{S}'_i)$, and $c_i(\bm{S}_i,x_i)\geq c_i(\bm{S}'_i,x_i)$ for all $x_i\in X_i(\bm{S}_i)$.  
	
	\clearpage
	\begin{theorem}[\cite{Hoo17}] \label{th:merge}
		If the following conditions are satisfied, the merger of nodes with states $\bm{S}_i$ and $\bm{T}_i$ within a decision diagram results in a valid relaxation of the diagram.
		\begin{itemize}
			\item $\bm{S}_i\oplus \bm{T}_i$ relaxes both $\bm{S}_i$ and $\bm{T}_i$.  
			\item If state $\bm{S}'_i$ relaxes state $\bm{S}_i$, then given any control $v$ that is feasible in $\bm{S}_i$, $\phi(\bm{S}'_i,v)$ relaxes $\phi(\bm{S}_i,v)$. 
		\end{itemize}
	\end{theorem}
	
	All relaxed diagrams we consider are associated with a dynamic programming model that satisfies the conditions of the theorem.  The shortest path problem in the relaxed DD requires a modification of the original dynamic programming model that accounts for the merger of states.  Also, it is sometimes necessary to use additional state variables to obtain a valid relaxation \cite{BerCirHoeHoo14}, and so we replace the state vector $\bm{S}_i$ with a possibly enlarged vector $\bar{\bm{S}}_i$.  The recursive model becomes 
	\begin{equation}
	{\ds
		\bar{h}_i(\bar{\bm{S}}_i) = \min_{x_i\in X_i(\bar{\bm{S}}_i)} \Big\{ c_i(\bar{\bm{S}}_i,x_i) + \bar{h}_{i+1}\Big(\rho_{i+1}\big(\phi_i(\bar{\bm{S}}_i,x_i)\big)\Big)\Big\}
	}
	\label{eq:recursionRelax}
	\end{equation}
	where $\rho_{i+1}(\bar{\bm{S}}_{i+1})$ is a relaxation of state $\bar{\bm{S}}_{i+1}$ that reflects the merger of states in stage $i+1$ of the recursion.  In the example, $\bar{\bm{S}}_i=\bm{S}_i$, since no additional states are necessary to formulate the relaxation.
	
	It will be convenient to distinguish {\em exact} from {\em relaxed} state variables in model (\ref{eq:recursionRelax}).  A state variable $S_{ij}$ in (\ref{eq:recursionRelax}) is exact if any sequence of controls $x_1,\ldots,x_{i-1}$ that leads to a given value of $S_{ij}$ in the original recursion (\ref{eq:recursion}) leads to that same value in the relaxed recursion (\ref{eq:recursionRelax}).  
	Otherwise $S_{ij}$ is relaxed.  In the example, neither $V_i$ nor $t_i$ is exact if any pair of states can be merged.  However, if we permit the merger of $(V_i,t_i)$ and $(V'_i,t'_i)$ only when $t_i=t'_i$, then state variable $t_i$ is exact.  
	In general we have the following, which is easy to show.
	
	\begin{lemma} \label{le:exact}
		A state variable $S_{ij}$ is exact if states $\bm{S}_i$, $\bm{S}'_i$ are merged only when $S_{ij}=S'_{ij}$.
	\end{lemma}

	\section{Lagrangian Duality and Decision Diagrams}
	
	Consider an optimization problem
	\begin{equation}
	z^* = \min_{\bm{x}\in\mathcal{X}} \big\{f(\bm{x}) \; \big| \; \bm{g}(x)=\bm{0}\big\} 
	\label{eq:l1}
	\end{equation}
	where $\bm{x}=(x_1,\ldots, x_n)$, $\bm{g}(\bm{x})=(g_1(\bm{x}), \ldots, g_m(\bm{x}))$ and $\bm{0}=(0,\ldots,0)$. The condition that $\bm{x}\in\mathcal{X}$ is typically represented by a constraint set.  A {\em Lagrangian relaxation} of (\ref{eq:l1}) has the form
	\begin{equation}
	\theta(\bm{\lambda}) = \min_{\bm{x}\in\mathcal{X}} \big\{f(\bm{x}) + \bm{\lambda}^T\bm{g}(\bm{x})\big\}
	\label{eq:l2}
	\end{equation}
	where $\bm{\lambda}=(\lambda_1,\ldots,\lambda_m)$.  The relaxation {\em dualizes} the constraints $\bm{g}(\bm{x})=\bm{0}$.  It is easy to show that $\theta(\bm{\lambda})$ is a lower bound on $z^*$ for any $\bm{\lambda}\in\mathbb{R}^m$.  The {\em Lagrangian dual} of (\ref{eq:l1}) seeks the tightest bound $\theta(\bm{\lambda})$:
	\begin{equation}
	\max_{ \bm{\lambda}\in \mathbb{R}^m}\big\{\theta(\bm{\lambda})\big\}
	\label{eq:l3}
	\end{equation}
	The motivation for using a Lagrangian dual is to obtain tight bonds while dualizing troublesome constraints.  
	If $\bm{g}(\bm{x})$ depends on a very small number of state variables, then dualizing the constraint $\bm{g}(\bm{x})=0$ may allow one to solve the problem within time and space constraints. 
	
	Lagrangian duality can be illustrated in the minimum tardiness job sequencing problem discussed earlier.  The variables $x_1,\ldots,x_n$ should have different values, or equivalently, that each job $j$ should occur in a given solution exactly once.  Since a relaxed DD does not enforce this condition, we dualize the constraint $\bm{g}(\bm{x})=0$, where $\bm{g}(\bm{x})=(g_1(\bm{x}),\ldots,g_n(\bm{x}))$ and 
	\[
	g_j(\bm{x}) = -1 + \sum_{i=1}^n (x_i=j), \;\;\mbox{with} \;\;
	(x_i=j)=\left\{
	\begin{array}{ll} 1 & \mbox{if $x_i=j$} \\ 0 & \mbox{otherwise}
	\end{array}
	\right.
	\]
	As observed in \cite{BerCirHoe15}, the Lagrangian penalty $\lambda_j g_j(\bm{x})$ can be represented in a relaxed DD by adding $\lambda_j$ to the cost of each arc corresponding to control $j$ and subtracting $\sum_j \lambda_j$ from the cost of each arc leaving the root node.  Then the length of each $r$--$t$ path includes the Lagrangian penalty $\bm{\lambda}^T\bm{g}(\bm{x})$ for the corresponding solution $\bm{x}$, which is zero if $\bm{x}$ satisfies the all-different constraint.
	
	
	In general, $\bm{g}(\bm{x})$ can be computed recursively in a relaxed DD when there is a vector-valued {\em immediate penalty function} $\bm{\gamma}_i(\bar{\bm{S}}^L_i,x_i)$ for which
	\begin{equation}
	\bm{g}(\bm{x}) = \sum_{i=1}^n \bm{\gamma}_i(\bar{\bm{S}}_i^L,x_i)
	\label{eq:penalty}
	\end{equation}
	where 
	each $\bar{\bm{S}}_i^L$ is a tuple consisting of {\em exact} state variables $S_{i\ell}$ for $\ell\in L$.  
	In the example, the constraint function $\bm{g}(\bm{x})$ requires no state information, and $\bar{\bm{S}}_i^L$ is empty.  The immediate penalty 
	$\bm{\gamma}_i((V_i,t_i),x_i)$ can be written simply $\bm{\gamma}_i(x_i)$, where
	\begin{equation}
	\gamma_{ij}(x_i) = \left\{
	\begin{array}{l@{\hspace{2ex}}l}
	(x_i=j) & \mbox{if $i\in\{2, \ldots, n\}$} \\
	(x_i=j) -1 & \mbox{if $i=1$}
	\end{array}
	\right.
	\label{eq:dualize}
	\end{equation}
	for $j=1,\ldots,n$.  
	
	Implementing a Lagrangian relaxation (\ref{eq:l1}) in a relaxed DD also requires that the original objective function value $f(\bm{x})$ be computed as part of the path length.  Normally, the path length is only a lower bound on $f(\bm{x})$.  To compute $f(\bm{x})$ exactly in the relaxed recursion (\ref{eq:recursionRelax}), we must have an immediate cost function that depends only on exact state variables.  That is, we must have 
	\begin{equation}
	f(\bm{x}) = \sum_{i=1}^n \bar{c}_i(\bar{\bm{S}}_i^K,x_i)
	\label{eq:cost}
	\end{equation}
	where $\bar{\bm{S}}_i^K$ is a tuple consisting of exact state variables $S_{ik}$ for $k\in K$.  In the example, the immediate cost function (\ref{eq:dp5}) depends on the state variable $t_i$, which must therefore be exact.  We therefore have $\bar{\bm{S}}_i^K=(t_i)$.  Due to Lemma~\ref{le:exact}, we can ensure that $t_i$ is exact by merging nodes only when $t_i$ has the same value in the corresponding states.  Thus Lagrangian relaxation can be implemented in the minimum tardiness example if we merge nodes in this fashion.  
	
	The above observations can be summed up as follows.
	
	\begin{theorem} \label{th:L}
		Lagrangian relaxation (\ref{eq:l2}) can be implemented in a relaxed DD if there are immediate cost functions $\bar{c}_i(\bar{\bm{S}}_i^K,x_i)$ for which (\ref{eq:cost}) holds and immediate penalty functions $\bm{\gamma}_i(\bar{\bm{S}}_i^L,x_i)$ for which (\ref{eq:penalty}) holds, where $\bar{\bm{S}}_i^K$ and $\bar{\bm{S}}_i^L$ consist entirely of exact state variables in $\bar{\bm{S}}_i$.  In this case the recursion for computing shortest paths in the relaxed DD becomes 
		\[
		{\ds
			\bar{h}_i(\bar{\bm{S}}_i,\bm{\lambda}) = \min_{x_i\in X_i(\bar{\bm{S}}_i)} \Big\{ \bar{c}_i(\bar{\bm{S}}_i^K,x_i) + \bm{\lambda}^T\bm{\gamma}_i(\bar{\bm{S}}_i^L,x_i) + \bar{h}_{i+1}\Big(\rho_{i+1}\big(\phi_i(\bar{\bm{S}}_i,x_i)\big),\bm{\lambda}\Big)\Big\}
		}
		\label{eq:recursionRelaxL}
		\]
	\end{theorem}
	
	\begin{corollary}
		Lagrangian relaxation (\ref{eq:l2}) can be implemented in a relaxed DD if nodes are merged only when their states agree on the values of the state variables on which the immediate cost functions and the immediate penalty functions depend.  That is, nodes with states $\bar{\bm{S}}_i$ and $\bar{\bm{T}}_i$ are merged only when $\bar{\bm{S}}_i^K=\bar{\bm{T}}_i^K$ and $\bar{\bm{S}}_i^L=\bar{\bm{T}}_i^L$, where $K$ and $L$ are in as Theorem~\ref{th:L}.
	\end{corollary}

	\section{Problem Classes}
	
	We now examine a few classes of job sequencing problems to determine whether they are suitable for Lagrangian relaxation on a relaxed DD.  All of these problems have an all-different constraint that is dualized as before using the immediate penalty function (\ref{eq:dualize}).  Since this function depends on no state variables, the state variables that must be exact are simply those on which the immediate cost function depends.  That is, $\bar{\bm{S}}_i^L$ is empty, and suitability for relaxation depends on which variables are in $\bar{\bm{S}}_i^K$.  We note that even when DD-based Lagrangian relaxation is not suitable for a given problem class, it may be useful when combined with branching, or when the relaxed DD is embedded in a larger model.

	\subsection{Sequencing with Time Windows}
	
	Problems in which jobs with state-independent processing times are sequenced, possibly subject to time windows, are generally conducive to Lagrangian relaxation on DDs.  The problem of minimizing total tardiness is discussed above, and computational results are presented in Section~\ref{computational}.  Minimizing makespan or the number of late jobs is treated similarly.  
	
	A popular variation on the problem minimizes the sum of penalized earliness and tardiness with respect to a common due date \cite{BisFel01,BisFel05,Che96,HallPos91,HallPosSet91,OwMor89,YinLinLu17}.  Earliness of job $j$ is weighted by $\alpha_j$ and lateness by $\beta_j$.  The recursive model for an exact DD uses the same state variables $(V_i,t_i)$ as the minimum tardiness problem.  However, a valid relaxed DD requires an additional state variable $s_i$ that represents latest start time, while $t_i$ again represents earliest finish time.  The transition and immediate cost functions are
	\[
	\begin{array}{l}
	{\ds
		\bar{\phi}_i\big((V_i,s_i,t_i),x_i\big) = \big(V_i\cup\{x_i\},\;s_i+p_{x_i},\;t_i+p_{x_i}\big)
	} \vs \\
	\bar{c}_i\big((V_i,s_i,t_i),x_i\big) = \alpha_{x_i}\big(s_i+p_{x_i}-d_{x_i}\big)^+ + \beta_{x_i}\big(t_i+p_{x_i}-d_{x_i}\big)^+
	\end{array}
	\]
	The merger operation is
	\[
	(V_i,s_i,t_i)\oplus (V'_i,s'_i,t'_i) = \big(V_i\cap V'_i, \max\{s_i,s'_i\},\min\{t_i,t'_i\}\big)
	\]
	The immediate cost depends on both $s_i$ and $t_i$, which means that both of these state variables must be exact.  This may appear to result in a large relaxed DD, because nodes can be merged only when they agree on both state variables.  However, since $s_i$ and $t_i$ are initially equal, and these state variables are exact, they remain equal throughout the relaxed DD construction.  The resulting DD is therefore the same that would result if a single state variable were exact.  Computational results are presented Section~\ref{computational}.

	\subsection{Time-Dependent Costs and/or Processing Times}
	
	Costs and processing times can be time-dependent in two senses: they may depend on the position of each job in the sequence, or on the clock time at which the job is processed.  Both senses occur in the literature, and both can be treated with Lagrangian relaxation on DDs.  
	
	If the processing time $p_{ij}$ of job $j$ depends on the position $i$ of the job in the sequence, then the immediate cost in the relaxation is 
	\[
	\bar{c}_i((V_i,t_i),x_i) = \big(\max\{t_i,r_{x_i}\} + p_{ix_i} - d_{x_i}\big)^+
	\]
	which depends only on the state variable $t_i$.  Since $\bar{c}_i$ is already indexed by the position $i$, any other element of cost that depends on $i$ is easily incorporated into the function. Thus we need only ensure that states are merged only when they agree on $t_i$, a condition that is already satisfied in the relaxed model described above for minimum-tardiness sequencing problems.
	
	If the processing time $p_j(s)$ of job $j$ depends on the time $s$ at which job $j$ starts, the immediate cost is
	\[
	\bar{c}_i((V_i,t_i),x_i) = \big(\max\{t_i,r_{x_i}\} + p_{x_i}(\max\{t_i,r_{x_i}\}) - d_{x_i}\big)^+
	\]
	which again depends only on state variable $t_i$.  Any other time-dependent element of cost likewise depends only on $t_i$, and so states can be merged whenever they agree on $t_i$.

\subsection{Sequence-Dependent Processing Times}
	
We refer to a job $j$'s processing time as sequence dependent when its processing time $p_{j'j}$ depends on the immediately preceding job $j'$ in the sequence.  When there are no time windows and the objective is to minimize travel time, the problem is a traveling salesman problem.  The state variables are $V_i$ and the immediately preceding job $y_i$.  The transition and immediate cost functions are
\begin{equation}
	\begin{array}{l}
	{\ds
		\bar{\phi}_i\big((V_i,y_i),x_i\big) = \big(V_i\cup\{x_i\},x_i\big)
	} \vs \\
	{\ds
		\bar{c}_i\big((V_i,y_i),x_i\big) = p_{y_ix_i}
	} 
	\end{array}
	\label{eq:seqDep}
\end{equation}
Since the immediate cost depends only on state $y_i$, nodes can be merged whenever they are reached using the same control.  This permits a great deal of reduction in the relaxed DD and suggests that a Lagrangian approach to bounding can be effective.  While pure traveling salesman problems are already well solved, a DD-based Lagrangian bounding technique may be useful when there are side constraints.  
	
When there are time windows in the problem, an additional state variable $t_i$ representing the finish time of the previous job is necessary for a stand-alone relaxed DD.  The transition function is 
\begin{equation}
	\bar{\phi}_i\big((V_i,y_i,t_i),x_i\big) = \big(V_i\cup\{x_i\},x_i,\max\{r_{x_i},t_i\}+p_{y_ix_i}\big)
	\label{eq:seqDep2}
\end{equation}
The immediate cost functions for minimizing travel time and total tardiness, respectively, are
\begin{align}
& \bar{c}_i\big((V_i,y_i,t_i),x_i\big) = (r_{x_i}-t_i)^+ + p_{y_ix_i}
	\label{eq:seqDep3} \\
& \bar{c}_i\big((V_i,y_i,t_i),x_i\big) = \big(\max\{r_{x_i},t_i\}+p_{y_ix_i} - d_{x_i}\big)^+ \label{eq:seqDep4}
\end{align}
In either case, the immediate cost depends on two state variables $y_i$ and $t_i$, and nodes can be merged only when they agree on these variables.  This is likely to result in an impracticably large relaxed DD.  For example, if there are 50 jobs and a few hundred possible values of $t_i$, a layer of the relaxed DD could easily expand to tens of thousands of nodes.  We confirmed this with preliminary experiments on the Dumas instances \cite{DumDesGelSol95}.  Lagrangian relaxation on a stand-alone relaxed DD therefore does not appear to be a promising approach to bounding TSP problems with time windows. One can, of course, use the simpler DD described by (\ref{eq:seqDep}) to bound travel time (although not total tardiness), as is done in \cite{BerCirHoe15}.  However, this relaxation ignores time windows altogether and would yield a weaker bound than (\ref{eq:seqDep2})--(\ref{eq:seqDep3}).

	\subsection{State-Dependent Processing Times}
	
	We refer to a job's processing times as state dependent when they depend on one or more of the state variables in the recursion, such as the set $V_i$ of jobs already processed.  Such a problem is studied in \cite{Hoo17}, where the processing time is less if a certain job has already been processed.  State variables are again $V_i$ and $t_i$, but to build a relaxed DD we need an additional state variable $U_i$ representing the sets of jobs that have been processed along some path to the current node.  The transition function and immediate cost function are
	\[
	\begin{array}{ll}
	{\ds 
		\bar{\phi}_i\big((V_i,U_i,t_i),x_i\big) = \big(V_i\cup\{x_i\},U_i\cup\{x_i\},\max\{r_{x_i},t_i\}+p_{x_i}(U_i)\big)
	} \vs \\
	{\ds
		\bar{c}_i\big((V_i,U_i,t_i),x_i\big) = \big(\max\{r_{x_i},t_i\}+p_{x_i}(U_i) - d_{x_i}\big)^+
	}
	\end{array}
	\]
	where the processing time is $p_{x_i}(U_i)$.  The merger operation is
	\[
	(V_i,U_i,t_i) \oplus (V'_i,U'_i,t'_i) = \big(V_i\cap V'_i, U_i\cup U'_i, \min\{t_i,t'_i\}\big)
	\]
	Since the cost depends on both $t_i$ and $U_i$, these state variables must be exact, and states can be merged only when they agree on the values of $t_i$ and $U_i$.  This predicts that the relaxed DD will grow rapidly with the number of jobs.  \mbox{DD-based} Lagrangian relaxation is therefore not a promising approach to this type of problem.

	\section{Computational Experiments} \label{computational}
	
	\subsection{Problem Instances}
	
	To assess the quality of bounds obtained from Lagrangian relaxation on relaxed DDs, it is necessary to obtain problem instances with known optimal values, or values that are likely to be close to the optimum.  We carry out tests on two well-known sets of instances, corresponding to two sequencing problems identified earlier to be suitable for bounding.  One is the set of minimum weighted tardiness instances of Crauwels, Potts and Wassenhove \cite{CraPotWas98}, which we refer to as the CPW instances.  The other is the Biskup-Feldman collection of minimum weighted earliness-plus-tardiness instances with a common due date \cite{BisFel01}.
	
	The CPW set consists of 125 instances of each of three sizes: 40 jobs, 50 jobs, and 100 jobs.  We compute bounds for first 25 instances in the 40- and 50-job sets.  These instances exhibit a wide range of gradually increasing tardiness values, thus providing a diverse selection for testing.  Optimal solutions are given in \cite{BisFel01} for all of these instances except instance 14 with 40 jobs, and instances 11, 12, 14 and 19 with 50 jobs.  Solution values presented in \cite{BisFel01} for the unsolved instances are apparently the best known.
	
	The Biskup-Feldman collection includes 10 instances of each size, where the sizes are 10, 20, 50, 100 and 200 jobs.  We study the instances with 20, 50 and 100 jobs.  The instances specify only the processing times $p_j$ and the earliness/tardiness weights $\alpha_j$, $\beta_j$ described earlier.  The common due date for all jobs in an instance is not specified.  Typical practice is to set the due date equal to $d(h) = \lfloor h \sum_j p_j\rfloor$, where $h$ is a parameter.  
	
	We compare our lower bounds against the best known solution values reported in \cite{YinLinLu17}.  These authors compute the earliness  penalty with respect to $d(h_1)$ and the tardiness penalty with respect to $d(h_2)$ for $h_1<h_2$, so that the penalty for each job $j$ is
	$\alpha_j(d(h_1)-t_j)^+ + \beta_j(t_j-d(h_2))^+$, where $t_j$ is the finish time of job $j$.  Heuristics are used in \cite{YinLinLu17} to solve instances with $(h_1,h_2)=(0.1,0.2)$, $(0.1,0.3)$, $(0.2,0.5)$, $(0.3,0.4)$, and $(0.3,0.5)$.  We study instances with $(h_1,h_2)=(0.1,0.2)$ and $(0.2,0.5)$ to provide a look at contrasting cases.  To our knowledge, none of these instances have been solved to proven optimality.

	\subsection{Solving the Lagrangian Dual}
	
	We use subgradient optimization to solve the Lagrangian dual problem (\ref{eq:l3}).  Each iterate $\bm{\lambda}^k$ is obtained from the previous by the update formula
	\[
	\bm{\lambda}^{k+1} = \bm{\lambda}^k + \sigma_k \bm{g}(\bm{x}^k)
	\]
	where $\bm{x}^k$ is the value of $\bm{x}$ obtained when computing $\theta(\bm{\lambda}^k)$ in (\ref{eq:l2}).  
	The art of Lagrangian optimization is choosing the step size $\sigma_k$.  This choice is avoided in \cite{BerCirHoe15} by using the Kelly-Cheney-Goldstein bundle method \cite{Lem01} of deriving $\bm{\lambda}^k$ from previous iterates.  However, Polyak's method \cite{Pol87} seems better suited to our purposes, because it is much easier to implement, and it requires as a parameter only an upper bound $\theta^*$ on the optimal value of $\theta(\bm{\lambda})$.  Such a  bound is available in practice, because one seeks to estimate how far a known solution value lies from the optimal value, and that solution value is an upper bound $\theta^*$.  Polyak's method defines the step size to be
	\[
	\sigma_k = \frac{\theta^* - \theta(\bm{\lambda}^k)}{\|\bm{g}(\bm{x}^k)\|^2_2}
	\]

	\subsection{Building the Relaxed Diagram}
	
	In previous work \cite{AndHadHooTie07,BerCirHoeHoo13,BerCirHoeHoo16}, heuristically-selected nodes are merged in each layer after all states obtainable from the previous layer are generated.  Since we wish to merge all nodes that agree on $t_i$, and no others, we merge these nodes as we generate the states, rather than first generating all possible states and then merging nodes.  This drastically reduces computation time and results in reasonable widths that gradually increase as layers are created.  

	\subsection{Computational Results}
	
	The computational tests were run on a Dell XPS-13 laptop computer with Intel Core i7-6560U (4M cache, 3.2 GHz) and 16GB memory.  Results for the CPW instances appear in Table~\ref{ta:CPW}.  The table displays optimal (or best known) values and DD-based bounds, as well as the absolute and relative gap between the two.  It also shows the maximum width of the relaxed DD (always obtained in layer $n$), the time required to build the DD, and the time consumed by the subgradient algorithm.  Since a subgradient iteration requires only the solution of a shortest path problem in the relaxed DD, we allowed the algorithm to run for 50,000 iterations to obtain as much improvement in the bound as seemed reasonably possible.  Due to slow convergence, which is typical for subgradient algorithms, a bound that is nearly as tight can be obtained by executing only, say, 20\% as many iterations.  
	
	The bounds in Table~\ref{ta:CPW} are reasonably tight.  The gap is well below one percent in most cases, and below 0.1\% in about a quarter of the cases, although a few of the bounds are rather weak.  The optimal value was obtained for one instance.  The gap for instance 14 in the 40-job table suggests that the best known value is probably not optimal, while no such inference can be drawn for the 4 unsolved 50-job instances.
	
	Results for the Biskup-Feldman instances appear in Table~\ref{ta:BF}, where the bounds are compared with the best known solutions.  The relaxed DDs are the same for the two sets of due dates $(h_1,h_2)=(0.1,0.2), (0.2,0.5)$; only the costs differ.  The bounds are very tight, resulting in gaps that are mostly under 0.1\%.  This indicates that the known solutions are, at worst, very close to optimality.  In fact, optimality is proved for 8 instances, which represents the first time that any of these instances have been solved.  The bounds may be equal to the optimal value for other instances, since the known values displayed may not be optimal.

	\section{Conclusion}
	\vspace{-1ex}
	
	We have shown how Lagrangian relaxation in a stand-alone relaxed decision diagram can yield tight optimization bounds for certain job sequencing problems.  We also characterized problems on which this approach is likely to be effective; namely, problems in which a relaxed DD of reasonable width results from a restricted form of state merger. The restriction is that states may be merged only when they agree on the values of state variables on which the cost function and dualized constraints depend in a recursive formulation of the problem.

	
	Based on this analysis, we observed that job sequencing problems with state-independent processing times and time windows are suitable for this type of bounding, whether one minimizes tardiness, makespan or the number of late jobs.  The same is true when processing times are dependent on when the job is processed or its position in the sequence.  The traveling salesman problem can also bounded in this fashion.  However, the TSP with time windows, as well as problems with state-dependent processing times in general, are normally unsuitable for DD-based Lagrangian relaxation, unless the relaxed diagram is combined with branching or embedded in a larger model.  
	
	We ran computational experiments on 110 instances from the well-known Crauwels-Potts-Wassenhove and Biskup-Feldman problem sets, with sizes ranging from 20 to 100 jobs.  
	We found that DD-based Lagrangian relaxation can provide tight bounds for nearly all of these instances.  This is especially true of the Biskup-Feldman instances tested, all of which were unsolved prior to this work.  We showed that the best known solutions are almost always within a small fraction of one percent of the optimum, and we proved optimality for 8 of the solutions.  To our knowledge, these are the first useful bounds that have been obtained for these instances.

	More generally, our analysis can be used to identify dynamic programming models that may have a useful relaxation based on relaxed decision diagrams and Lagrangian duality.

	\begin{landscape}
		
		\begin{table}
			\fontsize{8}{10}\selectfont
			\centering
			\caption{Comparison of bounds with optimal values (target) of CPW instances. Computation times are in seconds.}
			\label{ta:CPW}
			\begin{tabular}{c@{\hspace{5ex}}c}
				\begin{tabular}{r@{\hspace{2ex}}r@{\hspace{2ex}}r@{\hspace{2ex}}r@{\hspace{2ex}}r@{\hspace{3ex}}r@{\hspace{2ex}}r@{\hspace{2ex}}r}
					\multicolumn{8}{c}{40 jobs} \\ [0.5ex]
					Instance & Target & Bound & Gap & Percent & Max   & Build & Subgr \\
					&        &       &     & gap     & width & time  & time \\ [0.5ex]
					\hline \\ [-2ex]
					1  &    913 &    883 &  30 &  3.29\% & 3163 & 16 & 1287 \\ 
					2  &   1225 &   1179 &  46 &  3.76\% & 3652 & 20 & 1420 \\
					3  &    537 &    483 &  54 & 10.06\% & 3556 & 20 & 1443 \\
					4  &   2094 &   2047 &  47 &  2.24\% & 3568 & 20 & 1427 \\
					5  &    990 &    980 &  10 &  1.01\% & 3305 & 18 & 1312 \\
					6  &   6955 &   6939 &  16 &  0.23\% & 3588 & 20 & 1406 \\
					7  &   6324 &   6299 &  25 &  0.40\% & 3509 & 20 & 1437 \\
					8  &   6865 &   6743 & 122 &  1.78\% & 3508 & 20 & 1393 \\
					9  &  16225 &  16049 & 176 &  1.08\% & 3699 & 22 & 1468 \\
					10 &   9737 &   9591 & 146 &  1.50\% & 3426 & 19 & 1346 \\
					11 &  17465 &  17417 &  48 &  0.27\% & 3770 & 23 & 1493 \\
					12 &  19312 &  19245 &  67 &  0.35\% & 3644 & 22 & 1435 \\
					13 &  29256 &  29003 & 253 &  0.86\% & 3736 & 22 & 1506 \\
					14 & $^*$14377 
					&  14100 & 277 &  1.93\% & 3609 & 21 & 1406 \\
					15 &  26914 &  26755 & 159 &  0.59\% & 3849 & 23 & 1554 \\
					16 &  72317 &  72120 & 197 &  0.27\% & 3418 & 19 & 1382 \\
					17 &  78623 &  78501 & 122 &  0.16\% & 3531 & 20 & 1384 \\
					18 &  74310 &  74131 & 179 &  0.24\% & 3524 & 20 & 1431 \\
					19 &  77122 &  77083 &  39 &  0.05\% & 3407 & 19 & 1320 \\
					20 &  63229 &  63217 &  12 &  0.02\% & 3506 & 20 & 1344 \\
					21 &  77774 &  77754 &  20 &  0.03\% & 3766 & 22 & 1433 \\
					22 & 100484 & 100456 &  28 &  0.03\% & 3489 & 20 & 1382 \\
					23 & 135618 & 135617 &   1 & 0.001\% & 3581 & 21 & 1375 \\
					24 & 119947 & 119914 &  33 &  0.03\% & 3477 & 19 & 1295 \\
					25 & 128747 & 128705 &  42 &  0.03\% & 3597 & 22 & 1339 \\
					\hline
					\multicolumn{8}{l}{\hspace{1ex} $^*$Best known solution}
				\end{tabular}
				&
				\begin{tabular}{r@{\hspace{2ex}}r@{\hspace{2ex}}r@{\hspace{2ex}}r@{\hspace{2ex}}r@{\hspace{3ex}}r@{\hspace{2ex}}r@{\hspace{2ex}}r}
					\multicolumn{8}{c}{50 jobs} \\ [0.5ex]
					Instance & Target & Bound & Gap & Percent & Max   & Build & Subgr \\
					&        &       &     & gap     & width & time  & time \\ [0.5ex]
					\hline \\ [-2ex]
					1  &   2134 &   2100 &  34 &  1.59\% & 4525 & 48 & 2633 \\ 
					2  &   1996 &   1864 & 132 &  6.61\% & 4453 & 53 & 2856 \\
					3  &   2583 &   2552 &  31 &  1.20\% & 4703 & 52 & 2697 \\
					4  &   2691 &   2673 &  18 &  0.67\% & 4585 & 55 & 2703 \\
					5  &   1518 &   1342 & 176 & 11.59\% & 4590 & 52 & 2658 \\
					6  &  26276 &  26054 & 222 &  0.84\% & 4490 & 48 & 2562 \\
					7  &  11403 &  11128 & 275 &  2.41\% & 4357 & 45 & 2499 \\
					8  &   8499 &   8490 &   9 &  0.11\% & 4396 & 46 & 2501 \\
					9  &   9884 &   9507 & 377 &  3.81\% & 4696 & 52 & 2660 \\
					10 &  10655 &  10594 &  61 &  0.57\% & 4740 & 53 & 2738 \\
					11 &$^*$43504 &  43472 &  32 &  0.07\% & 4597 & 50 & 2606 \\
					12 &$^*$36378 &  36303 &  75 &  0.21\% & 4500 & 48 & 2655 \\
					13 &  45383 &  45310 &  73 &  0.16\% & 4352 & 47 & 2521 \\
					14 &$^*$51785 &  51702 &  83 &  0.16\% & 4699	& 52 & 2656 \\
					15 &  38934 &  38910 &  47 &  0.12\% & 4650 & 52 & 2630 \\
					16 &  87902 &  87512 & 390 &  0.44\% & 4589 & 49 & 2623 \\
					17 &  84260 &  84066 & 194 &  0.23\% & 4359 & 45 & 2526 \\
					18 & 104795 & 104633 & 162 &  0.15\% & 4448 & 47 & 2505 \\
					19 &$^*$89299 &  89163 & 136 &  0.15\% & 4609 & 50 & 2660 \\
					20 &  72316	&  72222 &  94 &  0.13\% & 4678 & 51 & 2659 \\
					21 & 214546	& 214476 &  70 &  0.03\% & 4406 & 47 & 2580 \\
					22 & 150800	& 150800 &   0 &     0\% & 4098 & 39 &  418 \\
					23 & 224025	& 223922 & 103 &  0.05\% & 4288 & 44 & 2441 \\
					24 & 116015	& 115990 &  25 &  0.02\% & 4547 & 49 & 2620 \\
					25 & 240179	& 240172 &   7 &  0.003\% &4639 & 51 & 2686 \\
					\hline
					\multicolumn{8}{l}{\hspace{1ex} $^*$Best known solution}
				\end{tabular}\\ \vspace{-2.0ex}
			\end{tabular}
		\end{table}
		
		\clearpage

		\begin{table}
			\fontsize{8}{9}\selectfont
			\centering
			\caption{Comparison of bounds with best known values (target) of Biskup-Feldman instances.} \label{ta:BF}
			\begin{tabular}{r@{\hspace{1ex}}|@{\hspace{1ex}}r@{\hspace{2ex}}r@{\hspace{2ex}}r@{\hspace{2ex}}r@{\hspace{2ex}}r@{\hspace{2ex}}|@{\hspace{1ex}}r@{\hspace{2ex}}r@{\hspace{2ex}}r@{\hspace{2ex}}r@{\hspace{2ex}}r@{\hspace{1ex}}|@{\hspace{1ex}}r@{\hspace{2ex}}r}
				& \multicolumn{5}{c}{$(h_1,h_2)=(0.1,0.2)$} & \multicolumn{5}{c}{$(h_1,h_2)=(0.2,0.5)$} & & \\ [0.5ex]
				Instance & Target & Bound & Gap & Percent & Subgr & Target & Bound & Gap & Percent & Subgr & Max & Build \\
				&        &       &     & gap     & time  &        &       &     & gap     & time  & width & time \\ [0.5ex]
				\hline 
				20 jobs & & & & & & & & & & & & \\
				1  &4089& 4089& 0& 0\%    & 1   &1162&1162& 0&   0\%& 1&   323 & 0.12 \\
				2  &8251& 8244& 7& 0.08\% &28   &2770&2766& 4&0.14\%&27&   287 & 0.08 \\
				3  &5881& 5877& 4& 0.07\% &27   &1675&1669& 6&0.36\%&28&   287 & 0.08 \\
				4  &8977& 8971& 6& 0.07\% &27   &3113&3108& 5&0.16\%&27&   287 & 0.08 \\
				5  &4028& 4024& 4& 0.10\% &32   &1192&1187& 5&0.42\%&32&   341 & 0.10 \\
				6  &6306& 6288&18& 0.29\% &26   &1557&1557& 0&   0\%& 1&   271 & 0.09 \\
				7 &10204&10204& 0& 0\%    &1    &3573&3569& 4&0.11\%&29&   305 & 0.09 \\
				8  &3742& 3739& 3& 0.08\% &25   & 990& 979&11&1.11\%&25&   267 & 0.08 \\
				9  &3317& 3310& 7& 0.21\% &21   &1056&1055& 1&0.09\%&22&   230 & 0.07 \\
				10 &4673& 4669& 4& 0.09\% &29   &1355&1349& 6&0.44\%&30&   320 & 0.09 \\
				\hline
				50 jobs & & & & & & & & & & & & \\
				1  &39250&39250& 0&   0\% & 16  &12754&12752& 2& 0.02\%&501&   931 & 2.8 \\
				2  &29043&29043& 0&   0\% &191  & 8468& 8463& 5& 0.06\%&524&   931 & 2.9 \\
				3  &33180&33180& 0&   0\% &300  & 9935& 9935& 0&    0\%& 66&   836 & 2.4 \\
				4  &25856&25847& 9&0.03\% &549  & 7373& 7335&38& 0.52\%&521&   932 & 2.8 \\
				5  &31456&31439&17&0.05\% &540  & 8947& 8938& 9& 0.10\%&529&   932 & 3.0 \\
				6  &33452&33444& 8&0.02\% &544  &10221&10213& 8& 0.08\%&532&   932 & 2.9 \\
				7  &42234&42228& 6&0.01\% &491  &12002&11981&21& 0.17\%&465&   835 & 2.4 \\
				8  &42218&42203&15&0.04\% &491  &11154&11141&13& 0.12\%&478&   833 & 2.4 \\
				9  &33222&33218& 4&0.01\% &503  &10968&10965& 3& 0.03\%&508&   884 & 2.7 \\
				10 &31492&31481&11&0.03\% &529  & 9652& 9650& 3& 0.03\%&522&   932 & 2.9 \\
				\hline
				100 jobs & & & & & & & & & & & & \\
				1  &139573 &139556& 17& 0.01\% &4075 & 39495& 39467&28&0.07\%&3968&1882	&42	\\
				2  &120484 &120465& 19& 0.02\% &4065 & 35293& 35266&27&0.08\%&4068&1882	&44	\\
				3  &124325 &124289& 36& 0.03\% &3957 & 38174& 38150&24&0.06\%&4059&1882	&42	\\
				4  &122901 &122876& 25& 0.02\% &3903 & 35498& 35467&31&0.09\%&3964&1882	&42	\\
				5  &119115 &119101& 14& 0.01\% &3925 & 34860& 34826&34&0.10\%&4016&1882	&42	\\
				6  &133545 &133536&  9&0.007\% &3987 & 35146& 35123&23&0.07\%&3961&1882	&43	\\
				7  &129849 &129830& 19& 0.01\% &4027 & 39336& 39303&33&0.08\%&3974&1882	&43	\\
				8  &153965 &153958&  7&0.005\% &3722 & 44963& 44927&36&0.08\%&3865&1784	&39	\\
				9  &111474 &111466&  8&0.007\% &3930 & 31270& 31231&39&0.12\%&4008&1882	&42	\\
				10 &112799 &112792&  7&0.006\% &3936 & 34068& 34048&20&0.06\%&4003&1882	&42	\\
				\hline
			\end{tabular}\\ \vspace{-2.0ex}
		\end{table}
		
	\end{landscape}


\end{document}